# The measurement by scattering of the chiral order-parameter of spins in a crystal

Stephen W. Lovesey[1,2]

[1]ISIS Facility, STFC, Chilton, Oxfordshire OX11 0QX, UK
[2]Diamond Light Source Ltd., Oxfordshire OX11 0DE, UK

Fukunaga et al. [1] claim to measure in a crystal the chiral order-parameter of spins related to an expectation value of the product of spin operators, $\langle S^\alpha S^\beta \rangle$. Upon scrutiny, this claim appears false because the technique of neutron Bragg diffraction, which they use, does not reveal $\langle S^\alpha S^\beta \rangle$ or a proper chiral order-parameter. Instead, Bragg diffraction reveals the product of expectation values, $\langle S^\alpha \rangle \langle S^\beta \rangle$, an all together different and less informative quantity. This statement is true of both the intensity of a Bragg reflection and the polarization carried in the diffracted beam.

A chiral order-parameter in real space is a *non-local* property [2,3,4]. It is equivalent to information on at least three distinct positions in the medium under investigation. The mean value throughout the medium of the vector triple product of distinct positions is adequate to determine its chirality because $\mathbf{R} \cdot (\mathbf{R}' \times \mathbf{R}'')$ is both non-local and a scalar, or rather pseudo-scalar, quantity that is parity-odd and time-even.

In a motif of spins, where $\mathbf{S}_j$ is the spin operator at the jth. site in the crystal and $\mathbf{S}_j \times \mathbf{S}_j = i \mathbf{S}_j$, chirality is measured by a two-point operator $\mathbf{S}_i \times \mathbf{S}_j$ projected on to a third axis taken to be the relative distance between spins $\mathbf{R}_i - \mathbf{R}_j$. With Fourier components labelled by $\mathbf{k}$ of the spin density $\mathbf{S}(\mathbf{k}) = \Sigma_j \mathbf{S}_j \exp(i \mathbf{k} \cdot \mathbf{R}_j)$ a suitable chiral order-parameter is $i \langle \mathbf{\eta} \cdot \{\mathbf{S}(-\mathbf{k}) \times \mathbf{S}(\mathbf{k})\} \rangle$ in which $\langle ... \rangle$ denotes the expectation, or time-averaged, value of the enclosed quantum mechanical operator and $\mathbf{\eta} = \mathbf{k}/k$. Averaged over $\mathbf{\eta}$, the order parameter is proportional to the sum of $\langle \mathbf{d}_{ij} \cdot \{\mathbf{S}_i \times \mathbf{S}_j\} \rangle$ multiplied by $j_1(k|\mathbf{d}_{ij}|)$ with $\mathbf{d}_{ij} = \mathbf{R}_i - \mathbf{R}_j$ ($j_1(x)$ = spherical Bessel function of order 1). Our chiral order-parameter of spins, in common with the vector triple product of positions, is a time-even pseudo-scalar, and it can be different from zero in the absence of long-range magnetic order. In a scattering experiment, $\mathbf{k}$ is the change in the wave vector of the radiation upon deflection by the sample.

For Bragg diffraction, used by Fukunaga et al. [1], $\mathbf{k}$ must coincide with a vector in the reciprocal lattice of the crystal. Thus this diffraction arises from a perfect crystal created by averaging over all variables, including time, and it can only reveal entities that are quadratic forms of *local* quantities $\langle \mathbf{S}(\mathbf{k}) \rangle$ and the like, e.g., $\langle \mathbf{S}(-\mathbf{k}) \rangle \times \langle \mathbf{S}(\mathbf{k}) \rangle$ or a product of $\langle \mathbf{S}(\mathbf{k}) \rangle$ and the average density of nuclei. Certainly, Bragg diffraction can not reveal $\langle \mathbf{\eta} \cdot \{\mathbf{S}(-\mathbf{k}) \times \mathbf{S}(\mathbf{k})\} \rangle$.

To show that the latter can be measured in scattering it is best to look at Van Hove's formulation of the neutron cross-section [5] that employs a time-dependent spin correlation function $\langle S^\alpha(-\mathbf{k},0) \, S^\beta(\mathbf{k},t)\rangle$, where $\alpha$ and $\beta$ label Cartesian components of the spins and t is the time variable. A similar correlation function appears in the polarization of the scattered beam [6]. Consider, first, Bragg diffraction that appears in the limit $t \Rightarrow \infty$. In this limit $\langle S^\alpha(-\mathbf{k},0) \, S^\beta(\mathbf{k},t)\rangle \Rightarrow \langle S^\alpha(-\mathbf{k},0)\rangle \langle S^\beta(\mathbf{k},\infty)\rangle = \langle S^\alpha(-\mathbf{k},0)\rangle \langle S^\beta(\mathbf{k},0)\rangle = \langle S^\alpha(-\mathbf{k})\rangle \langle S^\beta(\mathbf{k})\rangle$ and we recover the finding used above for Bragg diffraction. Bragg diffraction can only be different from zero if there is magnetic order in the crystal, whereas the chiral order-parameter is time-even and possibly different from zero without violation of time-reversal symmetry.

The chiral order-parameter of spins is derived from the instantaneous value of the spin correlation function, namely, $\langle S^\alpha(-\mathbf{k},0) \, S^\beta(\mathbf{k},0)\rangle$ which is revealed in total scattering. The cross-section for the total scattering of energetic, polarized neutrons by a motif of spins contains a contribution $(\mathbf{\eta}\cdot\mathbf{P}) \, i \langle \mathbf{\eta}\cdot\{\mathbf{S}(-\mathbf{k}) \times \mathbf{S}(\mathbf{k})\}\rangle$ where **P** is polarization in the beam [6,7]. Since **P** and spin have the same symmetry properties it follows that the scalar product $(\mathbf{\eta}\cdot\mathbf{P})$ is a time-even pseudo-scalar, like photon helicity.

Circularly polarized x-rays may also reveal the chiral order-parameter [8]. The cross-section for the total scattering of energetic x-rays by a motif of spins, not surprisingly, contains $P_2 \, i \langle \mathbf{\eta}\cdot\{\mathbf{S}(-\mathbf{k}) \times \mathbf{S}(\mathbf{k})\}\rangle$ where $P_2$ is the photon helicity, or circular polarization. In a study of multiferroic terbium manganate by x-ray scattering, Fabrizi et al. [9] miss essential physics of the material with an examination limited to Bragg reflections.

Acknowledgements. I am grateful for useful discussions and correspondence with P. J. Brown, E. Balcar, L. C. Chapon, and S. P. Collins.